\documentclass[%
reprint,
preprintnumbers,
amsmath,amssymb,
aps,
pra,
longbibliography,
]{revtex4-1}

\usepackage{graphicx}
\usepackage{dcolumn}
\usepackage{bm}
\usepackage{hyperref}
\usepackage{color}
\usepackage{float}

\begin{document}

\title{%
Quadrupolar and dipolar excitons in symmetric trilayer heterostructures:\\Insights from first principles theory}

\author{Thorsten Deilmann}
  \affiliation{Institut f\"ur Festk\"orpertheorie, Universit\"at M\"unster, 48149 M\"unster, Germany}%
  \email{thorsten.deilmann@uni-muenster.de}
\author{Kristian Sommer Thygesen}
  \affiliation{CAMD, Department of Physics, Technical University of Denmark, DK-2800 Kongens Lyngby, Denmark}%

\date{June 4, 2024}

\begin{abstract}
Excitons in van der Waals heterostructures come in many different forms.
In bilayer structures, the electron and hole may be localized on the same layer or they may be separated forming an interlayer exciton with a finite out-of-plane dipole moment.
Using first principles calculations, we investigate the excitons in a symmetric WS$_2$/MoS$_2$/WS$_2$ heterostructure in the presence of a vertical electric field.
The excitons exhibit a quadratic Stark shift for low field strengths and a linear Stark shift for stronger fields.
This behaviour is traced to the coupling of interlayer excitons with opposite dipole moments, which lead to the formation of quadrupolar excitons at small fields.
The formation of quadrupolar excitons is determined by the relative size of the electric field-induced splitting of the dipolar excitons and the coupling between them given by the hole tunneling across the MoS$_2$ layer.
For the inverted structure, MoS$_2$/WS$_2$/MoS$_2$, the dipolar excitons are coupled by electron tunneling across the WS$_2$ layer.
Because this effect is much weaker, the resulting quadrupolar excitons are more fragile and break at a weaker electric field.
\end{abstract}

\maketitle

\section*{Introduction}
Van der Waals (vdW) heterostructures composed of vertically stacked two-dimensional (2D) materials \cite{Geim_2013} have ushered in a new paradigm for electronic band structure engineering and exciton physics.
Due to the weakness of the interlayer vdW interactions, the interfaces across a vdW heterostructure can be atomically sharp and the constituent 2D materials largely preserve their electronic properties.
This makes it possible to predict, with good accuracy, how the band structures of the individual 2D layers will line up when stacked.
By controlling the band alignment, different types of excitons can be realised.
Three main types of excitons have so far been established in vdW heterostructures,
namely intralayer excitons with the electron and hole residing in the same layer,
interlayer (IL) excitons with the electron and hole located on two different layers resulting in a finite out-of-plane dipole moment,
and hybridized versions of these \cite{OptExc2D,LightMatterHeteroRev}.

For heterobilayers with type-II band alignment, the lowest excitation is typically of the IL dipolar type \cite{kosmider_electronic_2013,rivera2015observation,Torun_2018,Calman_2020,WSe2CrI3}.
Due to their significant out-of-plane dipole moment carried by such IL excitons,
their energy can be tuned over more than 100 meV by means of a vertical electric field \cite{Mak_2018} (the linear Stark effect).
The negligible overlap of the electron and hole wave functions can lead to very long lifetimes of IL excitons \cite{palummo2015exciton},
which is advantageous for certain applications.
The downside is that their oscillator strength is so low that they become difficult to probe and control by optical methods.
If the energy of an IL exciton is close to that of an intralayer exciton,
a new type of exciton with mixed intralayer/interlayer character can form via hybridization (interlayer tunneling).
Such mixed IL excitons were theoretically predicted to exist in bilayer MoS$_2$ \cite{mixedIL}
and shown to combine the attractive properties of electrical tunability and finite oscillator strength.
These predictions were subsequently experimentally verified \cite{leisgang2020giant,peimyoo2021electrical,Lorchat_2021,Niehues_2019}.

By controlling the electronic band positions in a vdW heterostructure,
it is not only possible to manipulate the exciton energies and lifetimes,
but also the nature of the exciton-exciton interactions.
While intralayer excitons interact via an attractive exchange mechanism,
IL dipolar excitons interact via repulsive dipole-dipole interactions as has been demonstrated
by the redshift (blueshift) of the energy of intralayer (IL dipolar) biexcitons and higher exciton complexes \cite{li2020dipolar,kremser2020discrete}.
This opens up exciting possibilities for realising different types of interacting boson systems and quantum phases.

Very recently, excitons in symmetric trilayer (A/B/A) vdW heterostructures were measured
and interpreted as quadrupolar excitons \cite{Lian2023,Yu_2023,Li_2023,Bai_2023,xie2023bright}.
Such symmetric quadrupolar excitons have previously been studied theoretically and predicted to undergo a quantum phase transition
from a repulsive quadrupolar to an attractive staggered dipolar phase as the exciton density is increased \cite{slobodkin2020quantum,astrakharchik2021quantum}.
Depending on the energy of the intralayer excitons relative to the IL excitons and the strength of the interlayer tunneling
(both of which are largely governed by the band alignment)
intralayer excitons could also become relevant and further enrich the types of interactions and quantum phases that could be realised.

In this work, we establish a quantitative microscopic understanding of the factors governing the formation of quadrupolar excitons in symmetry transition metal dichalcogenide (TMDC) trilayers.
We carry out first principles many-body calculations of the excitonic spectrum
of the symmetric trilayer WS$_2$/MoS$_2$/WS$_2$ and its inverted counterpart MoS$_2$/WS$_2$/MoS$_2$ in the presence of a vertical electric field.
We focus on the electric field induced transition between quadrupolar and dipolar IL excitons and show that the critical field
at which the transition occurs differs significantly between the two structures as a result of the different coupling of the dipolar IL excitons across the middle layer.

\section*{The hetero-trilayer $\text{WS}_2$/$\text{MoS}_2$/$\text{WS}_2$}
\begin{figure}[t]
  \centering
  \includegraphics[width=.45\textwidth]{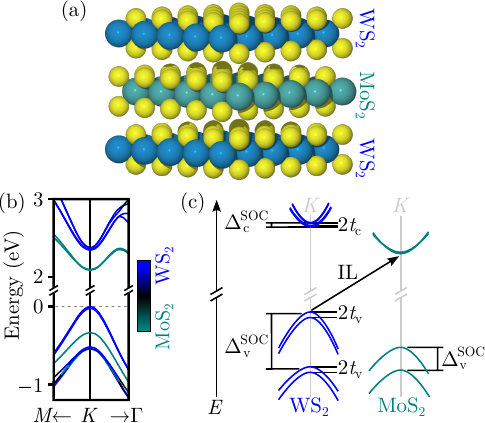}
  \caption{
    (a) Side view of the WS$_2$/MoS$_2$/WS$_2$  structure.
    (b) Calculated quasi-particle band structure of the WS$_2$/MoS$_2$/WS$_2$ trilayer close to the K point
    (zoom into the full band structure in the supporting Information).
    The color indicates the layer character.
    (c) Sketch with the bands localized on WS$_2$ (shown in blue)
    horizontally shifted from MoS$_2$ (in turquoise).
    Besides the splitting due to the spin-orbit coupling (SOC),
    the WS$_2$ bands are hybridized and split by $2t$.
    We underline that $\Delta^\text{SOC}$ and $t$ depend on the corresponding wave function
    and are much smaller for the conduction bands than the valence bands.
    The black arrow shows the energetically lowest interlayer transition (IL).
    Note that the energy is not shown to scale.
  }\label{fig1}
\end{figure}

We focus first on the hetero-trilayer WS$_2$/MoS$_2$/WS$_2$.
The three layer follow the $2H$ stacking of bulk TMDCs.
Adjacent layers are rotated by 180$^\circ$ (ABA stacking in Fig.~\ref{fig1}(a)),
which results in a mirror symmetry with respect to the central layer.
The similarity of the in-plane lattice constants of WS$_2$ (3.155\,\AA) and MoS$_2$ (3.160\,\AA) \cite{haastrup2018computational}
allows us to use a lattice-matched trilayer model, which renders $GW$ band structure and Bethe-Salpeter Equation (BSE) calculations feasible.
Note that the lattice-matched structure used here is fully consistent with the one used in our previous work on (mixed) IL excitons in MoS$_2$/WS$_2$ \cite{ILtrion}.

The band structure of the trilayer heterostructure is shown in Figures~\ref{fig1}(b,c).
Due to the hybridisation between the direct gap monolayers, the band gap of the trilayer becomes indirect with a size of about 1.9\,eV when evaluated in the $GW$ approximation
(see the supporting information for the complete band structure).
For the discussion of the optical properties,
the momentum direct transitions at $\pm$K are of main interest. The single-particle gap at $\pm$K is about $2.1$\,eV.

The trilayer WS$_2$/MoS$_2$/WS$_2$ forms two mirror symmetric type-II heterojunctions defined by the valence bands in the WS$_2$ layers and the conduction band in MoS$_2$.
Due to spin-orbit coupling (SOC), the valence bands (in all TMDCs) are split by more than 100\,meV at $\pm$K giving rise to two types of excitons usually referred to as A and B \cite{DarkExWSe2,molas_brightening_2017,echeverry_splitting_2016,DarkExTr}.
The effect of SOC on the conduction bands is significantly smaller, and gives rise to dark and bright excitons with almost same energy\cite{DarkExTr}.
In addition, the bands of WS$_2$ appear twice with slightly offset energies due to tunneling mediated by the MoS$_2$ layer.
Assuming a hopping matrix element of $t$, the band splitting is $2t$.

The size of the band splitting depends on the character of the wave function and is thus different for each band.
In particular, it is much smaller for the conduction bands compared to the valence bands.
The mediating middle layer also influences the hybridisation strength and band splitting.
Indeed, removing the MoS$_2$ layer makes the band splitting negligible.

As the trilayer is symmetric with a mirror symmetry in the central Mo layer,
each WS$_2$ band has the same weight (50\%) on each layer.
This symmetry can be broken by application of a vertical electrical field.
In the following we explore the effect of such a field on the energy and spatial structure of the lowest excitons of the trilayer.

To determine the excitons from first-principles,
we employ many-body perturbation theory in the $GW$+BSE approximation \cite{GdW,Rohlfing_eh,RevModPhys.74.601}.
The dominant peaks in absorption stem from the A and B intralayer excitations
in WS$_2$ and MoS$_2$ \cite{MoS2_Qiu,ramasubramaniam_large_2012,chernikov_exciton_2014,RevExTMDC}.
These excitons all have energies above 2\,eV (see supporting information).
Here, we focus on the IL excitons at lower energies. In general, the optical amplitude of IL excitons in vdW heterostructures are at least a factor of 50 smaller than those of the intralayer excitons.
However, it is still possible to observe them in photoluminescence spectra \cite{heteroMoWS2,nayak_probing_2017}.

\begin{figure*}[t]
  \centering
  \includegraphics[width=.8\textwidth]{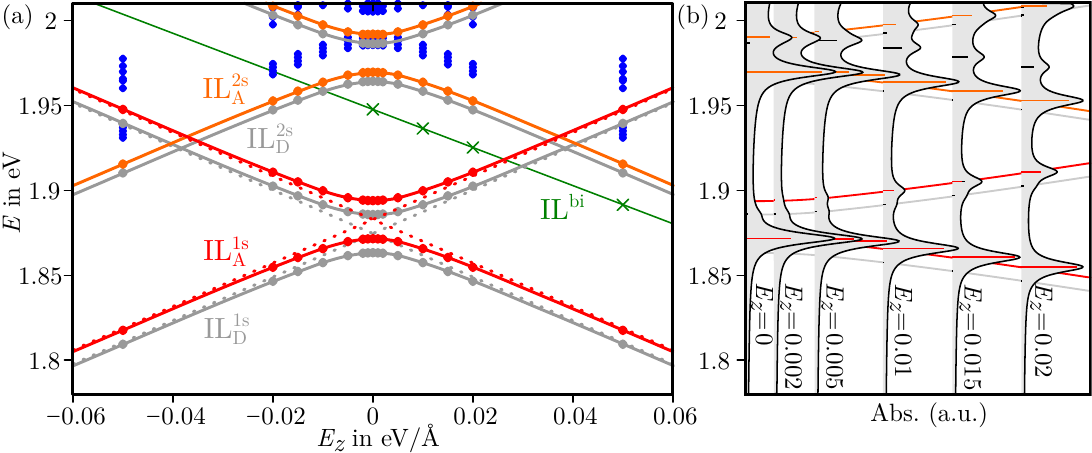}
  \caption{
    (a) The dots show the energy of interlayer excitons below the intralayer excitons (see Fig.~S2 for a larger energy scale) as function of a vertical electric field $E_z$.
    The full lines are a fits to the form $E^\text{X}\pm\sqrt{(p_\text{3L} E_z)^2 + t^2}$.
    Here $t$ and $p_{3L}$ are the matrix element coupling the two oppositely oriented dipolar excitons at the WS$_2$/MoS$_2$ and WS$_2$/MoS$_2$ interfaces,
    and the magnitude of their dipole moment.
    The expected energy of uncoupled dipolar excitons ($\pm p_\text{3L} E_z$) are shown by the dashed red lines.
    The colors denote the character of the excitons.
    The red IL$_A^\text{1s}$ are the ground state interlayer excitons,
    the orange IL$_A^\text{2s}$ are the first excited states.
    For both states an optically almost dark state is lying directly below.
    In blue the following excitons are shown (not discussed in detail).
    For comparison, the green crosses show the results of bilayer MoS$_2$/WS$_2$ \cite{ILtrion}.
    (b) Absorption spectra for the specified $E_z$ (in eV/\AA).
    The identified excitons from (a) are colored accordingly.
    We note that the oscillator strength is small and experimental identification requires specialized techniques \cite{Barr_2022}.
  }\label{fig2}
\end{figure*}

In Figure~\ref{fig2}(b) the optical absorption spectra are shown for different values of the vertical electric field.
In a type-II heterobilayer the IL excitons have a large dipole moment
and shift linearly with the electric field.
For MoS$_2$/WS$_2$ we find an energy shift of the form $p_\text{2L} E_z$ with an exciton dipole of $p_\text{2L} = 1.12/$\AA \cite{ILtrion}.
For the trilayer we observe a similar trend at large electric fields where the energy follows $\pm p_\text{3L} E_z$ with $p_\text{3L} = 1.28/$\AA.
The two branches with opposite field dependence originate from the mirror symmetric dipolar IL excitons with the hole in the left and right WS$_2$ layer, respectively.

In Figure~\ref{fig2}(a) the red curve shows the lowest excitation with non-vanishing optical amplitude.
Close to zero field strength the energies clearly deviate from the linear $E_z$-dependence.
By diagonalising a Hamiltonian for two initially degenerate dipolar excitons of energy $E^\text{X}$ and coupling $t$ in a vertical electric field,
\begin{equation}\label{eq:t}
  \left(\begin{array}{cc}
  E_\text{X} + p_\text{3L} E_z & -t \\
  -t & E_\text{X} - p_\text{3L} E_z
  \end{array}\right)
\end{equation}
we obtain energies of the form $E^\text{X} \pm \sqrt{(p_\text{3L} E_z)^2 + t^2}$ \cite{Yu_2023}.
At $E_z=0$ the splitting is $2t$ with $t=11.2$\,meV
(for the relation between $t$ and $t_v$/$t_c$ see the discussion below).
In addition to these IL$_\text{A}^\text{1s}$ excitons (red curve),
we find identical trends for the first excited IL$_\text{A}^\text{2s}$ excitons
(orange curve). The corresponding (almost) spin-dark states are shown in light gray. We note that these dark excitons have been observed experimentally in the monolayers \cite{xie2023bright,DarkExTr}.
In contrast to the monolayers, the bands of the different layers mix in the trilayer
and the lowest (almost dark) exciton IL$_\text{D}^\text{1s}$ gains oscillator strength for increasing $E_z$
(even though it remains small and is thus hardly visible).
Note that this increase agrees well with Xie et al. \cite{xie2023bright}
Above the IL$_\text{A}^\text{2s}$ excitons, other excitons are found (blue points).
These states belong to higher Rydberg states and are not that well converged in our calculations but seem to show a slightly reduced splitting $2t$.
In comparison to the dipolar IL excitons of the WS$_2$/MoS$_2$ bilayer, the corresponding exciton energies of the trilayer are slightly red shifted.
We find that this is a result of the decreasing band gap at $\pm$K of about $140$\,meV. This band gap reduction is similar to that found for MoS$_2$ and WS$_2$ homobilayers and homotrilayers of about $130$\,meV \cite{MoS2Dru,GdWTMDC}.

Returning to the optical absorption spectra in Figure~\ref{fig2}(b)
we find that the energetically lower branch has a larger oscillator strength.
This results from its bonding character, similar to mixed IL excitons in bilayers \cite{mixedIL}.
Furthermore, we note that while the oscillator strength of the intralayer 2s states (not seen on the plotting scale) are smaller compared to intralayer $1s$ states,
this is not the case for the IL 2s states, which may allow for their observation in optical experiments.

\begin{figure*}[t]
  \centering
  \includegraphics[width=.70\textwidth]{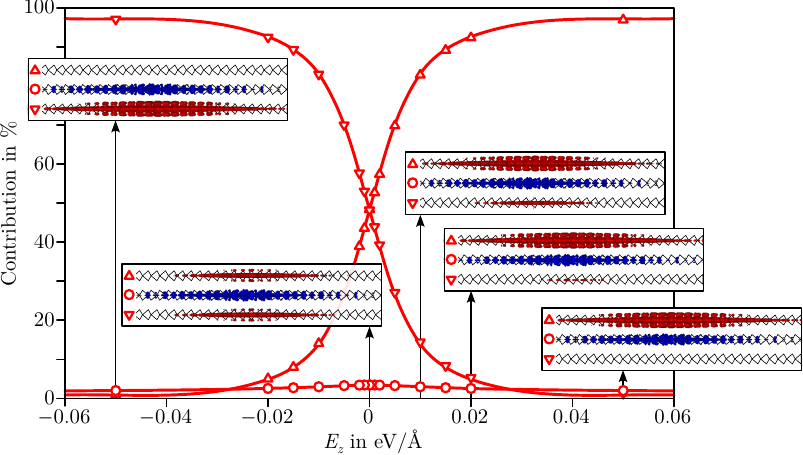}
  \caption{
    Analysis of the wave function of the lowest bright exciton (IL$_\text{A}^\text{1s}$) in the WS$_2$/MoS$_2$/WS$_2$ trilayer.
    The electron is fixed at the center of the MoS$_2$ layer and the percentage of hole on the upper WS$_2$ layer ($\Delta$),
    lower WS$_2$ layer ($\nabla$) and central MoS$_2$ layer ($\circ$) are shown as function of the vertical electric field.
    Lines are added as guide to the eye.
    The insets show a sideview of the hole component of the exciton wave function (red).
    The electron component is also shown (blue) with the hole fixed at the center of the lower WS$_2$ layer.
    A similar analysis for the dark IL$_\text{D}^\text{1s}$ exciton yields almost identical results.
  }\label{fig3}
\end{figure*}

The hybridization of the bands and their mixing close to $E_z = 0$
also have direct consequences for the spatial structure of the excitons.
Figure~\ref{fig3} shows the electron and hole distribution of the IL$_\text{A}^\text{1s}$ exciton on the different layers as function of $E_z$.
All hole distributions are evaluated for the electron fixed on an Mo atom of the central layer,
while the electron distributions are evaluated with the hole fixed on a W atom of the upper or lower layer.
\begin{figure*}[t]
  \centering
  \includegraphics[width=.74\textwidth]{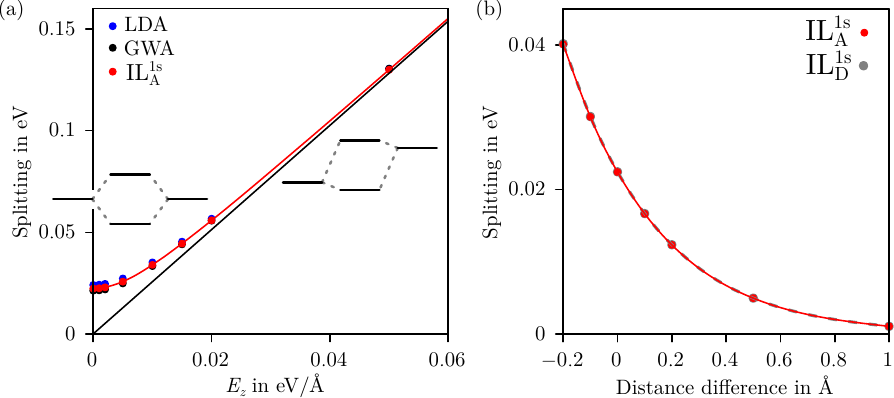}
  \caption{
    (a) Energy splitting of the two IL$_\text{A}^\text{1s}$ excitons (red),
    and the two uppermost valence bands (LDA in blue, $GW$ in black) in the trilayer WS$_2$/MoS$_2$/WS$_2$.
    The red curve shows a fit of the form $2\sqrt{(p_\text{3L} E_z)^2 + t^2}$ while the black line shows $2p_\text{3L}E_z$,
    i.e. the expected energy difference between uncoupled (oppositely oriented) dipolar excitons.
    The insets shows orbital coupling diagrams for $E_z=0$ and larger fields (Eq.~\ref{eq:t}).
    (b) The splitting ($2t$) for various different displacements of the layers. The zero point corresponds to the equilibrium layer distance as in (a).
    Both van der Waals gaps are changed by the same amount. The energy splitting of the bright and dark excitons IL$_\text{D}^\text{1s}$ (grey) are identical.
    Lines are shown as guide to the eye.
  }\label{fig4}
\end{figure*}
While for all $E_z$ the electron is localized on the central MoS$_2$ layer,
the contribution of the hole varies with $E_z$.
For large values of $E_z$ the hole is localized almost entirely on the upper layer.
Only a small part (less than 5\%) of the hole distribution is found on the central MoS$_2$ layer.
Therefore, in this regime the IL excitons resemble those of the heterobilayer \cite{ILtrion}.
At large negative fields the hole contribution is predominantly in the lower layer,
i.e. the dipole of the exciton as reversed.
For $E_z$ tending to zero, the hole becomes more equally distributed over the upper and lower layers.
Finally, for $E_z=0$ the mirror symmetry is restored and the hole is equally distributed on the upper and lower layer.
In conclusion, the application of the electric field can change the two quadrupolar excitons split by 2$t$ at $E_z=0$ into two oppositely oriented dipolar excitons.
We note in passing, that the small but roughly independent probability to find the hole on the middle layer
explains the small but non-vanishing optical amplitudes.

In Figure~\ref{fig4} we investigate the effects of the hybridization in more detail.
The red curve shows the splitting of the upper and lower branches of the
IL$_\text{A}^\text{1s}$ quadrupolar excitons in Figure~\ref{fig2}.
Again our first principles results nicely follow the trend
of the simple model $2\sqrt{(p_\text{3L} E_z)^2 + t^2}$.
Furthermore, we compare to the splitting of the upper valence bands (see Figure~\ref{fig1} and Figure~S1)
shown in black (GWA).
Only minor differences are seen,
the band splitting ($2t$ at $E_z=0$) $10.7$\,meV within the $GW$ approximation.
Including the electron-hole interactions yields a splitting of the two quadrupolar excitons of $11.2$\,meV discussed above.
Based on this we conclude that the exciton hybridization can be largely explained by the hole tunneling (valence band hybridisation).
This coupling results in a hybridization of the two dipolar excitons at small $E_z$.
We note in passing that already DFT-LDA (in blue) would yield a good approximation with $t=12.1$\,meV.

\begin{figure*}[t]
  \centering
  \includegraphics[width=.7\textwidth]{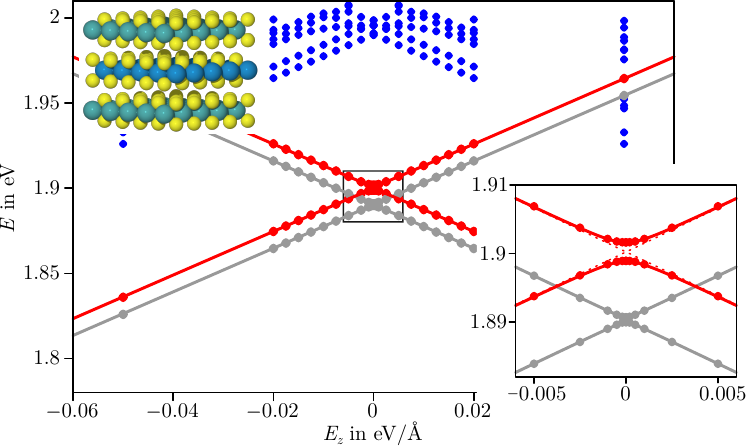}
  \caption{
    Excitation energies of MoS$_2$/WS$_2$/MoS$_2$ similar to Fig.~\ref{fig2}.
    The transition with optical amplitude is shown in red,
    an almost dark exciton is depicted in gray.
    All further excitations are blue.
    The inset at the top left is a 3D view.
    In addition, a zoom is shown close to zero for the framed box.
    Note that the mixing of the conduction bands is corrected for $0.0005$ and $0.001$\,eV/\AA.
  }\label{fig5}%
\end{figure*}%
In our calculations, the coupling of the layers can be varied by changing the interlayer layer distances (see Figure~\ref{fig4}(b)).
If we reduce the distance of the top and bottom layer by $0.2$\,\AA,
we find that the size of the splitting is almost doubled to about 40\,meV, i.e. $t=20$\,meV.
For an increased distance of $1$\,\AA\ per layer, the coupling is decreased to $0.5$\,meV.
This extreme sensitivity originates from the exponential decay of the WS$_2$ valence band wave functions (mediated by MoS$_2$) and underlines the necessity of very clean interfaces in experimental measurements.

\section*{Optical properties of $\text{MoS}_2$/$\text{WS}_2$/$\text{MoS}_2$}%
Another way to affect the coupling is via the choice of materials.
In Figure~\ref{fig5} we show the energy of the lowest excitons in the MoS$_2$/WS$_2$/MoS$_2$ trilayer.
The lowest optically active excitons are shown in red
and again show an energy dependence on the electric field of the form $\pm p_\text{3L} E_z$.
In this case, however, the linear regime extends to much smaller fields.
In the zoom-in, it can be seen that the excitons in this system also have an avoided crossing form
$E^\text{X} \pm \sqrt{(p_\text{3L} E_z)^2 + t^2}$.
However, the coupling $t$ is much smaller with values $0.4$ and $1.4$\,meV for spin-dark (grey) and bright states, respectively.
This coupling is an order of magnitude smaller compared to that of the WS$_2$/MoS$_2$/WS$_2$ trilayer.
This can be understood from the energies and character of the relevant bands (see Figure~\ref{fig1}(c)).
In MoS$_2$/WS$_2$/MoS$_2$ the coupling of the IL dipolar excitons is governed by the overlap of the
MoS$_2$ conduction bands, which is much weaker than the coupling of the WS$_2$ valence bands.
This is fully compatible with our previous findings for bilayer MoS$_2$ \cite{mixedIL}.
We note in passing that stronger couplings are observed for states at higher energies (only partially visible).
A further possibility to enhance the coupling may be pressure \cite{Steeger_2023}.
We stress that many of the qualitative results obtained in the present study of the vdW trilayers MoS$_2$/WS$_2$/MoS$_2$ and WS$_2$/MoS$_2$/WS$_2$ in the $2H$-stacking configurations,
should hold for other mirror-symmetric trilayer structures.
In particular this applies to the quadratic field dependence of the excitons around $E_z=0$.

In recent experimental works \cite{Lian2023,Yu_2023,Li_2023,Bai_2023,xie2023bright}
IL excitons in symmetric TMDC trilayer Moir\'{e} structures have been investigated.
The Moir\'{e} potential leads to spatially dependent band gaps, exciton energies, and layer hybridisation.
The description of such systems from first principles is very challenging \cite{Naik_2022}.
However, we expect that much of the results and mechanisms discussed here for the lattice matched trilayer structure carry over to Moir\'{e} structures to a good approximation,
if the sample keeps the mirror symmetry with respect to the central layer.
For instance, Yu et al. \cite{Yu_2023} measured the splitting of the quadrupolar excitons in WSe$_2$/WS$_2$/WSe$_2$
(determined by the interlayer hole tunneling) and obtained a coupling $t$ of about 12\,meV in good agreement with our results.

\section*{Conclusions}%
We have presented a first-principles study of the low energy optical properties
of the symmetric vdW trilayers WS$_2$/MoS$_2$/WS$_2$ and MoS$_2$/WS$_2$/MoS$_2$.
Below the intralayer excitons of the individual layers,
we find interlayer quadrupolar excitons due to the type-II heterojunctions.
These excitons can be seen as bonding/anti-bonding combinations of the dipolar excitons with oppositely oriented dipoles.
By applying an electric field the dipolar components of the quadrupolar excitons can be disentangled and controlled individually.
We have shown that the hybridization of the dipolar excitons is determined by the hybridisation of the single-particle valence/conduction band wave functions of the outermost layers and thus depends sensitively on the layer types of stacking order.
In our analysis we have focused on the lowest (1s) bright excitons.
However, our calculations unravel the existence of a series of other interlayer excitons with similar properties,
including a pair of (almost) dark 1s excitons just below the bright ones, and two pairs of bright and dark 2s excitons at higher energies.

\section*{Methods}%
A comprehensive explanation of our ab-initio framework has recently been published \cite{OptExc2D}.
Here, we briefly summarize the steps applied for the trilayers in this work.
For our calculations we neglect the small difference of the lattice constants
of the individual layers ($a^{\text{WS}_2}_\text{lat} = 3.155$\,\AA\ and $a_\text{lat}^{\text{MoS}_2} = 3.160$\,\AA)
and use the MoS$_2$ lattice constant in all calculations.
We note that the usage of the 0.2\% smaller WS$_2$ lattice constant only leads to small quantitative differences.
A first approximation of the electronic structure is calculated with DFT(LDA)
which is also the starting point for our many-body perturbation theory calculations.
The spin-orbit coupling is fully taken into account by working with spinor wave functions.
On top of this, we perform a $GW$ calculation to evaluate the band structure of the system.
We employ the $GdW$(LDA) approximation \cite{GdWTMDC}
\begin{equation}\label{eq:GdW}
  \hat{H}^\text{QP} \approx \hat{H}^\text{LDA} + \overbrace{iG\underbrace{(W - W_\text{metal})}_{dW}}^{\Delta \Sigma},
\end{equation}
which has previously shown a good accuracy for TMDC monolayers up to bulk.
Here $G$ is the Green function and $W$ is the screend Coulomb interaction.
The fictitious metallic screening reproduces the LDA
$V^\text{LDA}_\text{xc} \approx i G W_\text{metal}$ well for many systems.
The energies can finally be evaluated by
$E_{m\vec{k}}^\text{QP} = E_{m\vec{k}}^\text{LDA} + \langle \psi_{m\vec{k}}^\text{LDA} | \Delta\Sigma(E_{m\vec{k}}^\text{QP}) | \psi_{m\vec{k}}^\text{LDA} \rangle$.
To evaluate the optical properties we transform the Bethe-Salpeter equation (BSE)
to an electron-hole basis with the matrix elements
\begin{equation}\label{eq:BSEmatrix}
  (E_c-E_v)A^S_{vc} + \sum_{v'c'}K^{AA}_{vc,v'c'}(\Omega_S)A^S_{v'c'} = \Omega_S A^S_{vc} .
\end{equation}
Its diagonalization leads to the exciton energies $\Omega_S$ and their amplitudes $A^S_{vc}$.
The absorption is then calculated by the imaginary part of the dielectric function
\begin{equation}\label{eq:eps-opt}
  \varepsilon_2(\omega) = \frac{16\pi e^2}{\omega^2} \sum_S \left|\frac{\vec{A}}{|\vec{A}|} \langle 0 | \vec{v} | S \rangle\right|^2 \delta(\omega-\Omega_S),
\end{equation}
where $\langle 0 | \vec{v} | S \rangle$ are the velocity matrix elements \cite{RevModPhys.74.601,Rohlfing_eh}.
The exciton wave function can be evaluated by
\begin{equation}\label{eq:exwf}
\Phi^{S}(x_h,x_e) = \sum_{{v}{c}} A^{S}_{vc} \phi_{v}^\ast(x_h) \phi_{c}(x_e).
\end{equation}

For our DFT calculations we employ a basis of Gaussian orbitals with decay constants from $0.16$ to $2.5$\,$a_\text{B}^{-2}$.
For the evaluation of the two-point functions $P$, $\varepsilon$, and $W$
we use plane waves up to an energy cutoff of $2.5$\,Ry.
For the band structures, we use $12\times 12$ points in the first Brillouin zone.
This is increased to $30\times 30$ points for the optical properties from the BSE.

\section*{Acknowledgements}
T.D. acknowledges financial support from the Deutsche Forschungsgemeinschaft (DFG, German Research Foundation) through Project No. 426726249 (DE 2749/2-1 and DE 2749/2-2).
The authors gratefully acknowledge the Gauss Centre for Supercomputing e.V. (www.gauss-centre.eu) for funding this project by providing computing time through the John von Neumann Institute for Computing (NIC) on the GCS Supercomputer JUWELS \cite{JUWELS} at Jülich Supercomputing Centre (JSC).
K.S.T. acknowledges support from the Novo Nordisk Foundation Challenge Programme 2021: Smart nanomaterials for applications in life-science, BIOMAG Grant No. NNF21OC0066526.
K.S.T. is a Villum Investigator supported by VILLUM FONDEN (grant no. 37789).

%

\pagebreak\clearpage\onecolumngrid
\begin{center}\bf\large%
Supporting information\\
Quadrupolar and dipolar excitons in symmetric trilayer heterostructures:\\Insights from first principles theory
\end{center}
\setcounter{equation}{0}
\setcounter{figure}{0}
\setcounter{table}{0}
\setcounter{page}{1}
\makeatletter
\renewcommand{\theequation}{S\arabic{equation}}
\renewcommand{\thefigure}{S\arabic{figure}}

\begin{figure}[h]
  \centering
  \includegraphics[width=.5\textwidth]{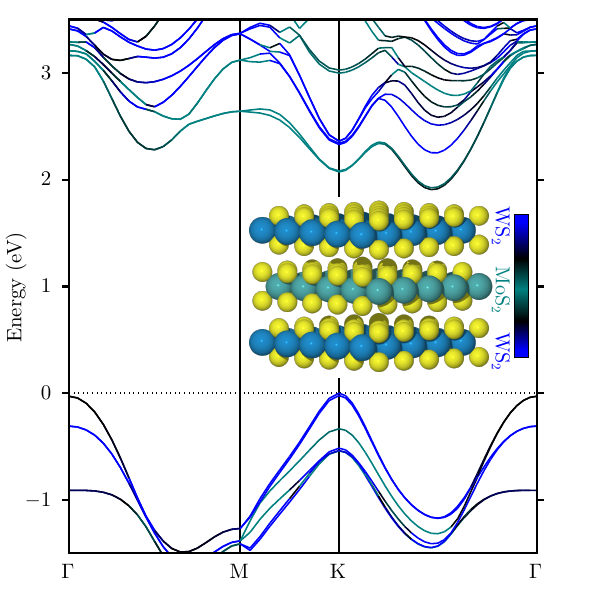}
  \caption{
    $GW$ band structure of WS$_2$/MoS$_2$/WS$_2$.
    The colors blue and turquoise show the contributions of the layers
    (compare inset).
  }\label{figS1}
\end{figure}
\begin{figure}[h]
  \centering
  \includegraphics[width=.7\textwidth]{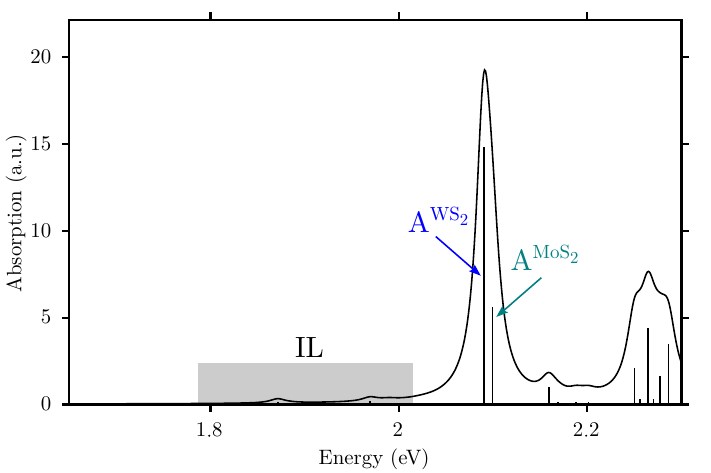}
  \caption{
    Optical absorption of WS$_2$/MoS$_2$/WS$_2$ from $GW$+BSE.
    The dominant peaks belong to the WS$_2$ and MoS$_2$ intralayer peaks.
    The gray region marks interlayer states which are discussed in the main text.
  }\label{figS2}
\end{figure}

\end{document}